\documentclass[12pt]{article}
\usepackage{graphicx}
\begin{document}

\noindent
{\bf The Pulsation Period of the Hot Hydrogen-Deficient Star MV Sgr}

\bigskip

\noindent
{\bf John R. Percy and Rong (Hannah) Fu\\Department of Astronomy and Astrophysics\\University of Toronto\\Toronto ON\\Canada M5S 3H4}

\bigskip

{\bf Abstract.}  MV Sgr is a hot, hydrogen-deficient star which has undergone
R CrB fadings.  We have used self-correlation analysis and Fourier analysis
of CCD V photometry in the AAVSO International Database
to identify a period of 8.0 days in this star; the amplitude is about 0.03 mag.
The variability is most likely due to pulsation.

\bigskip

\noindent
{\bf 1. Introduction}

\medskip

Hydrogen-deficient stars (Werner and Rauch 2008) are a rare but very diverse
(Jeffery 2008a) group of objects, in advanced and unusual stages of evolution.
In this paper, we are concerned with hydrogen-deficient stars which
can undergo the R CrB phenomenon -- unpredictable
fadings, followed by slow return to maximum brightness.  Most of the
approximately 50 such stars in our galaxy are cool -- ``classical" R CrB
stars -- but a few hot members of the group have been discovered.  

There are two proposed mechanisms for producing ``classical" R CrB stars:
the merger of a helium white dwarf and a carbon-oxygen white dwarf, and
a final helium shell flash.  Observational evidence seems to favor the
former mechanism, though a few R CrB stars may be produced by the latter
mechanism (Clayton 2011).  The hot hydrogen-deficient stars do not
necessarily arise from the same mechanism(s) as the cool ones, and may in
fact have diverse evolutionary histories (De Marco {\it et al.} 2002).

Many R CrB stars
are also pulsating variables, and the pulsation may be partly responsible
for the mass loss that leads to the fadings.  Pulsating hot hydrogen-deficient
variable stars have been classified as PV Tel stars, but Jeffery (2008b) argues that
this classification should be replaced by three new classes, based on the
pulsation period and mode in the star.

MV Sgr (AAVSO 1838-21, HV 4168, V $\sim$ 13.35) was discovered to be an R CrB star
in 1928 by Miss Ida Woods (Hoffleit 1959), and has been studied by various
techniques since then.  Its atmospheric properties are T$_{eff}$ = 16,000 $\pm$ 500 K and log g = 2.48 $\pm$ 0.30, and pulsations had not been found in MV Sgr as of 2008
(Jeffery 2008b).  

\medskip

\noindent
{\bf 2. Data, Analysis, and Results}

\medskip

Visual and CCD V data were taken from the American Association of Variable
Star Observers International Database (AID).  There were a total of approximately 2315 visual
observations, and 138 CCD V observations.  The former were made by 33 different
observers; the latter were made by
G. Di Scala, M. Simonsen, J. Temprano, and D. Wells.

The most numerous V observations were in the season JD 2455644-852.
Self-correlation analysis (Percy and Mohammed 2004) of these showed a
clear period of about 8.0 days, with a full amplitude of about 0.03 mag,
and at least 8 repeating minima, indicating coherent variability (Figure 1).  
Self-correlation anaysis of the whole
V dataset showed a period of 8.0 $\pm$ 0.1 days, with a similar amplitude.
The mean error of the observations, as determined from the intercept
on the vertical axis, is 0.03 mag.
Self-correlation analysis of
the visual data did not show a detectable signal, which is probably due to
the small amplitude and the much higher noise level.

The analysis was repeated with Fourier analysis, using the {\it Period04}
software (Lenz and Breger 2005).  For the V data: in the season JD 2455644-852,
and in the whole dataset,
the highest peaks were at frequencies of 0.128 cycle/day (period 7.8 days) and
and 0.122 cycle/day (period 8.2 days), respectively; the latter spectrum is
shown in Figure 2.  In each case: the highest
peak was only slightly higher than the next-highest peak.  It did, however,
agree with the period found from self-correlation analysis.  For the visual
data: the highest peak was at a frequency of 0.204 cycle/day (period 4.9 days),
but this was only slightly higher than the noise level, and may not be
significant, especially considering the small amplitude and the much higher
noise level in the visual data.  

\medskip

\noindent
{\bf 3. Discussion and Conclusions}

\medskip

MV Sgr displays a period of 8.0 days, which we assume to be due to pulsation.
The signal is not strong, but it is quite clear in the self-correlation
diagram, and is consistent with the results of the Fourier analysis.
This enables us to place it in Jeffery's (2008b) PV Tel I sub-class.

MV Sgr was not known to be pulsating.  We now have one more piece of useful
information about this star.  Also: the discovery of pulsation provides
further support for the possible connection between pulsation and the R CrB
phenomenon in hydrogen-deficient stars.

\begin{figure}
\includegraphics[height=8cm]{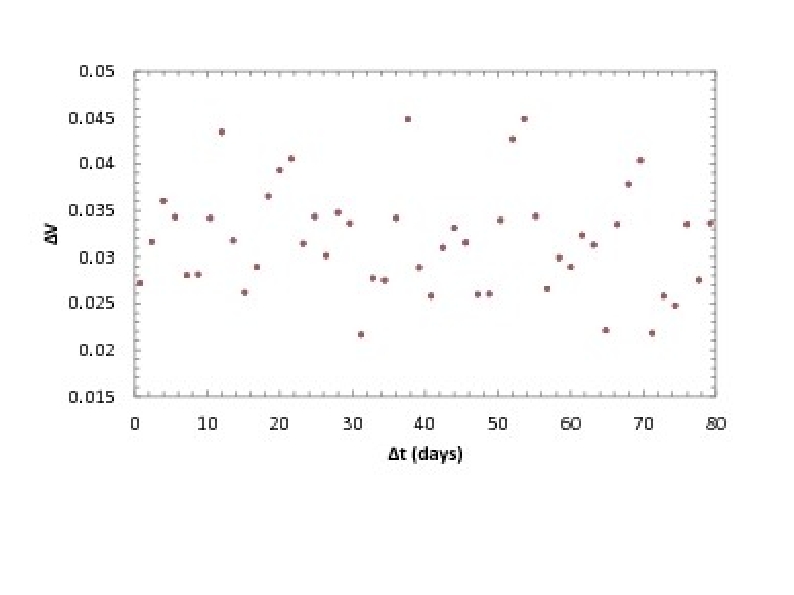}
\caption{Self-correlation diagram for CCD V observations of MV Sgr during the
season JD 2455644-852 when the observations are most numerous.  There are
repeating minima at multiples of 8.0 days: 8, 16, 24 (weak), 32, 40, 48, 56, 64 ...
days, and maxima at 4, 12, 20, 28 (weak), 36, 44, 52 ... indicating coherent variability.}
\end{figure}

\medskip

\begin{figure}
\includegraphics[height=8cm]{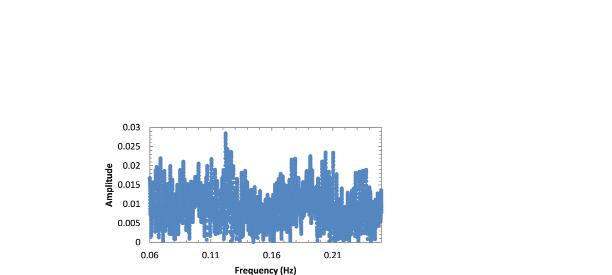}
\caption{Fourier spectrum for CCD V observations of MV Sgr.  The highest peak
is at a frequency corresponding to a period of 8.0 days.}
\end{figure}

\noindent
{\bf Acknowledgements}

\medskip

We thank the AAVSO observers and staff for making and archiving the visual
and CCD observations used in this study.  This study was supported by the
Ontario Work-Study
Program. 

\medskip

\noindent
{\bf References}

\smallskip

\noindent
Clayton, G.C. 2011, in {\it Asymmetric Planetary Nebulae V}, ed. A.A. Zijlstra {\it et al.}, Jodrell Bank Centre for Astrophysics, Manchester UK, 157.

\smallskip

\noindent
Hoffleit, D. 1959, {\it Astron. J.}, {\bf 64}, 241.

\smallskip

\noindent
Jeffery, C.S. 2008a, in {\it Hydrogen-Deficient Stars}, ed. K. Werner and
T. Rauch, ASP Conference Series, Vol. 391, 3.

\smallskip

\noindent
Jeffery, C.S. 2008b, {\it IAU Inf. Bull. Var. Stars}, \#5817.

\smallskip

\noindent
Lenz, P., and Breger, M. 2005, {\it Commun. Astroseismology}, {\bf 146}, 53.

\smallskip

\noindent
De Marco, O., {\it et al.} 2002, {\it Astron. J.}, {\bf 123}, 3387.

\smallskip

\noindent
Percy, J.R., and Mohammed, F. 2004, {\it J. Amer. Assoc. Var. Star Obs.}, {\bf 32}, 9.

\smallskip

\noindent
Werner, K. and Rauch, T. 2008 (editors), {\it Hydrogen-Deficient Stars},
ASP Conference Series, Vol. 391.

\end{document}